\newcommand{\tsecompldate}{1st May 2006, minor revisions 2nd August}
\documentclass[11pt]{article}

\usepackage{amsmath,amssymb,amscd}

\usepackage{graphicx}

\newcommand{\vol}[1]{\textbf{#1}}

\newcommand{\tpaptitle}[1]{``#1'',} 
\newcommand{\tpretitle}[1]{``#1'',} 
\newcommand{\tarttitle}[1]{``#1'',} 
\newcommand{\tbktitle}[1]{``#1''}     

\newcommand{\tref}[1]{(\ref{#1})}

\newcommand{\tversion}{Submitted version with minor corrections}

\newcommand{\tpre}[1]{} 
\newcommand{\tprenote}[1]{} 

\newcommand{\tnote}[1]{} 
\newcommand{\tcomment}[1]{} 
\newcommand{\tcommentx}[1]{} 

\newcommand{\href}[2]{#2}
\newcommand{\eprint}[1]{\texttt{#1}}

\newcommand{\tsedevelop}[1]{{}} 

\newcommand{\tsevec}[1]{\mathbf{#1}}
\newcommand{\tsemat}[1]{{\mathbf{\textsf{#1}}}}

\renewcommand{\tpre}[1]{#1}
\newcommand{\bea}{\begin{eqnarray}}
\newcommand{\eea}{\end{eqnarray}}
\newcommand{\beq}{\begin{equation}}
\newcommand{\eeq}{\end{equation}}
\newcommand{\nnel}{\nonumber \\ {}}

\newcommand{\evec}{\tsevec{e}}
\newcommand{\nvec}{\tsevec{n}}

\newcommand{\Mmat}{\tsemat{M}}

\newcommand{\Etilde}{\widetilde{E}}
\newcommand{\Ktilde}{\widetilde{K}}

\newcommand{\taverage}[1]{\langle #1 \rangle}
\newcommand{\kav}{\taverage{k}}

\begin{document}

\renewcommand{\thefootnote}{\fnsymbol{footnote}}

 \tpre{
 \begin{flushright}
 Accepted for inclusion at ECCS 2006 \\
 \texttt{Imperial/TP/06/TSE/3} \\
 \eprint{physics/0608052} \\
 \tsecompldate \\
 \tsedevelop{\tversion \\}
 \tsedevelop{(\texttt{rweccs.tex}  LaTeX-ed on \today ) \\}
 \end{flushright}
 \vspace*{1cm}
 }

\begin{center}
{\Large\textbf{Exact Solutions for Models of \\ Cultural
Transmission and Network Rewiring}}\tnote{tnotes such as this not
present in final
version} \\[0.5cm]
\tpre{\vspace*{1cm} }
 {\large T.S. Evans\footnote{WWW:
\href{http://www.imperial.ac.uk/people/t.evans}\texttt{http://www.imperial.ac.uk/people/t.evans}
}, A.D.K. Plato}
 \\[0.5cm]
 \tpre{\vspace*{1cm}}
 \href{http://www.imperial.ac.uk/research/theory}{Theoretical Physics},
 Blackett Laboratory, Imperial College London,\\
 London, SW7 2AZ,  U.K.
\end{center}

\begin{abstract}
We look at the evolution through rewiring of the degree distribution
of a network so the number edges is constant. This is exactly
equivalent to the evolution of probability distributions in models
of cultural transmission with drift and innovation, or models of
homogeneity in genes in the presence of mutation.  We show that the
mean field equations in the literature are incomplete and provide
the full equations.  We then give an exact solution for both their
long time solution and for their approach to equilibrium. Numerical
results show these are excellent approximations and confirm the
characteristic simple inverse power law distributions with a large
scale cutoff under certain conditions. The alternative is that we
reach a completely homogeneous solution. We consider how such
processes may arise in practice, using a recent Minority Game study
as an example.
\end{abstract}

\renewcommand{\thefootnote}{\arabic{footnote}}
\setcounter{footnote}{0}


\subsection*{Introduction}

The observation of power law probability distribution functions for
things as diverse as city sizes, word frequencies and scientific
paper citation rates has long fascinated people. Yule, Zipf, Simon
and Price \cite{Yule24,Zipf49,Simon55,Price65,Price76} provide some,
but by no means all, of the classic examples.  Coupled with modern
ideas of critical phenomena and self-organised criticality (see
\cite{Jensen98} for an introduction) this might suggest that such
power laws reflect fundamental, perhaps inviolable, processes behind
human behaviour.  These are the modern expositions of ideas that
have captivated for centuries as exemplified by Thomas
Hobbes\footnote{Thomas Hobbes (1588-1679) was a philosopher who held
that Human beings are physical objects, sophisticated machines whose
functions and activities can be described and explained in purely
mechanistic terms --- ``The universe is corporeal; all that is real
is material, and what is not material is not real''
\cite{Hobbes1651}.}.  So when power laws are mixed with modern icons
such as the World Wide Web \cite{BA99,BAJ99} we have an intoxicating
mixture.

In the context of complex networks, the focus is usually on power
laws in the degree distributions, the property which defines a
`scale-free' network (see \cite{Evans04} for a review and
references).  In growing networks, if one connects new vertices to
existing vertices chosen with a probability proportional to their
degree (at least this is the dominant behaviour for large degree)
--- `preferential attachment' \cite{BA99,BAJ99} --- then the degree
distribution for large degree $k$ is of the form $k^{-\gamma}$ with
a power $\gamma$ greater than two.

However, note that the degree distribution of a network is an
ultra-local property of its vertices.  The neighbours of a vertex at
the other end of the edges play no role, it does not matter that the
edge describes some bilateral relationship between two vertices.
This should not be surprising, the older studies such as Simon and
Price \cite{Simon55,Price65,Price76} make no
reference to a network.  One may be added easily to their examples
and models but it is not necessary.  Conversely, one need not refer
to the network of the World Wide Web, one can just count links on a
page.  For this reason the model of Simon \cite{Simon55} and that of
Barab\'{a}si, Albert and Jeong \cite{BA99,BAJ99} are identical
despite the fact the latter refer to a network, the former does not.
Thus it is only for convenience that in this paper we will use the
language of complex networks\footnote{There are some suggestions
about how such a power law might emerge \emph{only} because of the
network structure \cite{Vaz00,Vaz02,SK04,ES05}.}.  One may easily
dispense with the network as do many of our references.

We start by observing that there is much less material on the degree
distribution of networks which don't grow and their non-network
counterparts. The model of Watts and Strogatz \cite{WS98} is a
primary example of a non-growing network, but it does not produce a
power law. We will focus on models of non-growing networks where the
number of edges $E$ is constant (and non-network equivalents) with
power law degree distributions. The point is that these power laws
are invariably simple inverse powers of degree, $n(k) \propto 1/k$,
and so quite distinct from those found in most growing models.

Such non-growing models have also been shown to be relevant to a
wide range of examples. For instance it has been used when
considering the transmission of cultural artifacts: the decoration
of ceramics in archaeological finds \cite{Neiman95,BS03,BHS04}, the
popularity of first names \cite{HB03}, dog breed popularity
\cite{HBH04}, the distribution of family names in constant
populations \cite{ZM01}.  Similar models have been used to study the
diversity of genes \cite{KC64}.  The same types of probability
distribution have also been seen in a study of the Minority Game
\cite{ATBK04} and our model suggests why such features emerge even
when there is no explicit scale-free network. Such rewiring models
have also been studied for their own merits \cite{PLY05,XZW05}. For
definiteness in this paper we will use the language of cultural
transmission \cite{Neiman95,BS03,BHS04,HB03,HBH04}.

\subsection*{The Model}

\begin{figure}
{\centerline{\includegraphics[width=8cm]{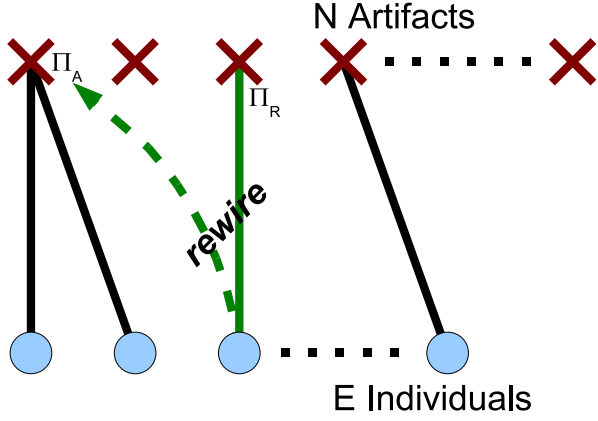}}}
\caption{Illustration of the model.} \label{fCopyModel}
\end{figure}

Consider a simple bipartite model\footnote{We have considered other
types of network and the generalisations are straightforward.} with
$E$ `individual' vertices, each with one edge\footnote{The degree of
the `individuals' does not effect the derivations and is only
relevant to the interpretation.} running to any one of $N$
`artifact' vertices, as shown in Figure~\ref{fCopyModel} .
The degree of the artifact vertices is $k$ indicating that one
artifact has been `chosen' by $k$ distinct individuals.  The
rewiring will be of the artifact ends of the edges, so each
individual is always connected to the same edge. It is the degree
distribution of the artifact vertices which we will consider so
$n(k)$ is the number of artifacts each of which has been chosen by
$k$ individuals.  The probability distribution of interest is then
$p(k) = n(k)/N$.

At each time step we make two choices but initially no changes to
the network.  First we choose an individual at random\footnote{We
adopt the common convention that `random' without further
qualification indicates that a uniform distribution is used to draw
from the set implicit from the context.} and consider its single
link. This is equivalent to choosing an edge at random. It is also
equivalent to picking an artifact vertex with pure preferential
attachment, that is with probability proportional to its degree. We
indicate the probability of choosing a particular artifact at this
stage as $\Pi_R$ since we are going to remove this edge from this
artifact.

The edge chosen is going to be attached to another artifact vertex
picked with probability $\Pi_A$.  This is the second choice we have
to make and it will be done with a mixture of preferential
attachment and random artifact vertex choice. In a fraction $p_r$ of
the attachment events we chose a random artifact vertex to receive
the rewired edge.  In the context of studies of cultural
transmission \cite{Neiman95,BS03,BHS04,HB03,HBH04} this corresponds
to \emph{innovation}, while in gene evolution it is \emph{mutation}
\cite{KC64}. Alternatively with probability $p_p$ we use
preferential attachment to find a new artifact vertex for
attachment.  This is \emph{copying} of the choice previously made by
another individual, \emph{drift} in the work on cultural
transmission \cite{Neiman95,BS03,BHS04,HB03,HBH04}, while it is the
inheritance mechanism in models of gene \cite{KC64} or family name
\cite{ZM01} homogeneity. If these are the only types of event
$p_p+p_r=1$, the number of artifacts $N$ is constant and
\beq
 \Pi_R = \frac{k}{E}, \qquad
 \Pi_A = p_r\frac{1}{N} + p_p\frac{k}{E},
   \qquad (0 \leq k \leq E) \; .
 \label{PiRPiAsimple}
\eeq
Note that there is a chance $\Pi_R\Pi_A$ that we will choose the
same artifact vertex for both attachment and removal and there will
then be no change in the network.

Finally, once both the artifacts for edge removal and addition have
been picked, we perform the rewiring.  The mean field equation for
evolution of $n(k)$ is then \cite{Evans06}
\begin{eqnarray}
\lefteqn{n(k,t+1) - n(k,t)}
 \nnel
 &=&   n(k+1,t) \Pi_R(k+1,t) \left( 1- \Pi_A(k+1) \right)
 \nnel
 &&
     - n(k,t)   \Pi_R(k,t)   \left( 1- \Pi_A(k)   \right)
     - n(k,t)   \Pi_A(k,t)   \left( 1- \Pi_R(k)   \right)
 \nnel
 &&
     + n(k-1,t) \Pi_A(k-1,t) \left( 1- \Pi_R(k-1) \right)
   ,
   \qquad (0 \leq k \leq E)
   \label{neqngen}
\end{eqnarray}
We must set $n(k)=\Pi_R(k)=\Pi_A(k)=0$ for $k=-1$ and $k=E+1$ to
ensure this equation is valid at the boundaries $k=0$ and $k=E$. It
is \emph{crucial} that we include the factors of $(1-\Pi_A)$ and
$(1-\Pi_R)$ otherwise the behaviour at the boundaries is incorrect.
We are explicitly excluding events where the same vertex is chosen
for removal and attachment in any one rewiring event as they do not
change the network but they are likely only if $n(k \sim E)=1$. Such
$(1-\Pi)$ terms are missing from other discussions of such models
but the literature usually has $n(k \sim E) \ll 1$ so these factors
are negligible. Thus the results in the literature will be
approximately the same as ours in this regime.

We can rephrase this as a Markov process.  Consider a vector
$\nvec(t)$ where $n_i(t) = n(E+1-i,t)$ for $i=1,2,\ldots,E+1$.  Then
we can think of the equations \tref{neqngen} as
\bea
 \nvec(t+1) &=& \Mmat \nvec(t).
 \eea
The transition matrix is
\bea
 \Mmat &:=&
 \begin{pmatrix}
g(E) & h(E-1) & 0 & \cdots &&0 &0 \\
f(E)&g(E-1) &h(E-2) & && & \\
0 & &\ddots & && &\vdots\\
0&&f(k+1)&g(k)&h(k-1)&&0\\
\vdots&&&\ddots&&&0\\
&&&&f(2)&g(1)&h(0)\\
0&0&&\cdots&0&f(1)&g(0)
 \end{pmatrix}
 .
 \label{transmatrix}
\eea
where the matrix entries are specified by the functions
\begin{subequations}
\begin{eqnarray}
f(k)&=&\Pi_R(k)(1-\Pi_A(k))\\
g(k)&=&1-f(k)-h(k)\\
h(k)&=&\Pi_A(k)(1-\Pi_R)
\end{eqnarray}
\end{subequations}
The evolution is then given by the eigenvectors and eigenvalues of
$\Mmat$\tnote{This matrix is guaranteed to have positive eigenvalues
with the largest equal to one??? Whose theorem is this? Frobenius
Perron?}
\bea
 \nvec(t) = c_1 \evec^{(1)} + \sum_{i=2}^{E+1} c_i \lambda_i^t \evec^{(i)},
 &&
 1 = \lambda_1 > \lambda_2 \geq \ldots \geq |\lambda_i| \geq \lambda_{i+1}
 \geq \ldots
 \label{tdepsol}
 \\
 \Mmat \evec^{(j)} = \lambda_j \evec^{(j)}
 &&
 j = 1,2,\ldots (E+1)
\eea

\subsection*{Stationary Solution}

The stationary solution for the degree distribution $n(k,t)=n(k)$,
the eigenvector associated with the largest eigenvalue
$\lambda_1=1$, can be found by substituting $n(k,t)=n(k)$ into the
evolution equation \tref{neqngen}. We then note that if the first
and third lines are equal then so are the second and third lines.
Thus we look for solutions of the form
\bea
 n(k+1) \Pi_R(k+1,t) \left( 1- \Pi_A(k+1) \right)
     = n(k)   \Pi_A(k,t)   \left( 1- \Pi_R(k)   \right) .
 \label{nstaticeqn}
\eea
The result is  \cite{Evans06}
\beq
 n(k) = A^{(1)}
 \frac{\Gamma( k + \frac{p_{r}}{p_p}\kav ) }
      {\Gamma (k+1 ) }
 \frac{\Gamma( \frac{E}{p_p} - \frac{p_{r}}{p_p}\kav  - k) }
      {\Gamma (E+1  -  k  ) }
 \qquad (E \geq k \geq 0)
 \label{nsimpsol}
\eeq
where $A$ is a constant normalisation and the average degree is
$\kav = E/N$.\tnote{Normalisation, total N and total E?}  This
solution has two characteristic parts. The first ratio of Gamma
functions for $E\gg k \gg 1$ behaves as
\beq
\frac{\Gamma( k + \frac{p_{r}}{p_p}\kav ) }
      {\Gamma (k+1 ) } \propto k^{-\gamma} \left( 1+ O( k^{-1} ) \right),
\qquad
 \gamma = 1- \frac{p_{r}}{p_p}\kav \leq 1 \; .
 \label{gammasimp}
\eeq
For $p_r E = Np_p$ we have an exact inverse power law.  The power is
always below one but for many values ($p_r \ll 1$) the power is
close to one. This is very different from the results for simple
models with growth in the number of edges where demanding that the
first moment is finite, $\kav<\infty$, requires $\gamma >2$.

However the $(1-\Pi_A)$ and $(1-\Pi_R)$ terms in \tref{neqngen} have
led to the second ratio of Gamma functions which if $p_p k \ll E$
gives an exponential cutoff
\bea
\frac{\Gamma\left( \frac{E}{p_p} - \frac{p_{r}}{p_p}\kav  - k\right)
} {\Gamma (E+1  -  k  ) }
     &\propto & \exp \{ -\zeta k \} ( 1+ O(\frac{k}{E}))
      , \qquad k \ll \left(\frac{E}{p_p} - \frac{p_{r}}{p_p}\right),
      \\
      \zeta &=& -\ln \left( p_p \right) -\frac{2 p_r \kav}{E}
      \label{zetasol}
            \\
      &\approx & p_r
      \mbox{ if }
      p_r \ll 1 , \;  \kav \ll E \; .
\eea
However the numerator of this second ratio of Gamma functions
becomes very large for $k=E$ if $(E-p_r \kav) \ll p_p$.  So this
happens if $p_r \ll p_{r, \mathrm{simple}}$, where
\beq
 p_{r,\mathrm{simple}} = \frac{1}{E+1-\kav} \approx \frac{1}{E} +
 \frac{(\kav-1)}{E^2} +O(E^{-2}) \; .
\eeq
This spike at $k=E$ will dominate the degree distribution. The point
where the distribution has become flat at the upper boundary, so
$n(E) = n(E-1)$  defines another critical random attachment
probability $p_{r ,\mathrm{crit}}$ at
\bea
 p_{r \mathrm{crit}} &=& \frac{E-1}{E^2+E(1-\kav)-1-\kav}
 \\
 E p_{r \mathrm{crit}} & \approx & 1+\frac{(\kav-2)}{E}
\eea
Thus when $p_r \lesssim 1/E$ the degree distribution will show a
spike at $k=E$.

Overall we see two distinct types of distribution. For large
innovation or mutation rates, $Ep_r \gtrsim 1$, we get a simple
inverse power with an exponential cutoff
\beq
 n(k) \propto (k)^{-\gamma} \exp\{-\zeta k\}
 ,
 \qquad
 p_r \gtrsim \frac{1}{E}
\eeq
This is the behaviour noted in the literature
\cite{PLY05,KC64,HBH04,BHS04,XZW05} and since $E p_r \gtrsim 1$ the
formulae of the literature for the power $\gamma$ and cutoff $\zeta$
are a good approximation to the exact formulae given here. Note that
in any one practical example it will be impossible to distinguish
the power law derived from the data from $\gamma=1$. The power
drifts away from one as we raise the innovation rate $p_r$ towards
one but only at the expense of the exponential regime starting at
lower and lower degree. That is, only when the power is very close
to one can we get enough of a power law to be significant.

However as $p_r$ is lowered towards zero we get a change of
behaviour in the exponential tail around $p_r E \approx 1$.  First
we find the exponential cutoff $\zeta^{-1}$ moves to larger and
larger values, eventually becoming bigger than $E$.  In fact this
second ratio of Gamma functions becomes equal to one for all $k$ at
$p_r=p_{r,\mathrm{simple}} \approx E^{-1}$.  At that value of
$p_r=p_{r,\mathrm{simple}}$ we have no cut off and we are closest to
an exact inverse power law for all degree values. Slightly below
that value of $p_r$ the tail starts to rise for $p_r >
p_{r,\mathrm{crit}} \approx E^{-1}$. For $p_r E \ll 1$, i.e.\ if
there has been no random artifact chosen after most edges have been
rewired once, then we will almost certainly find one artifact linked
to most of the individuals, $n(E) \approx 1$.

\begin{figure}
\includegraphics[width=7.0cm]{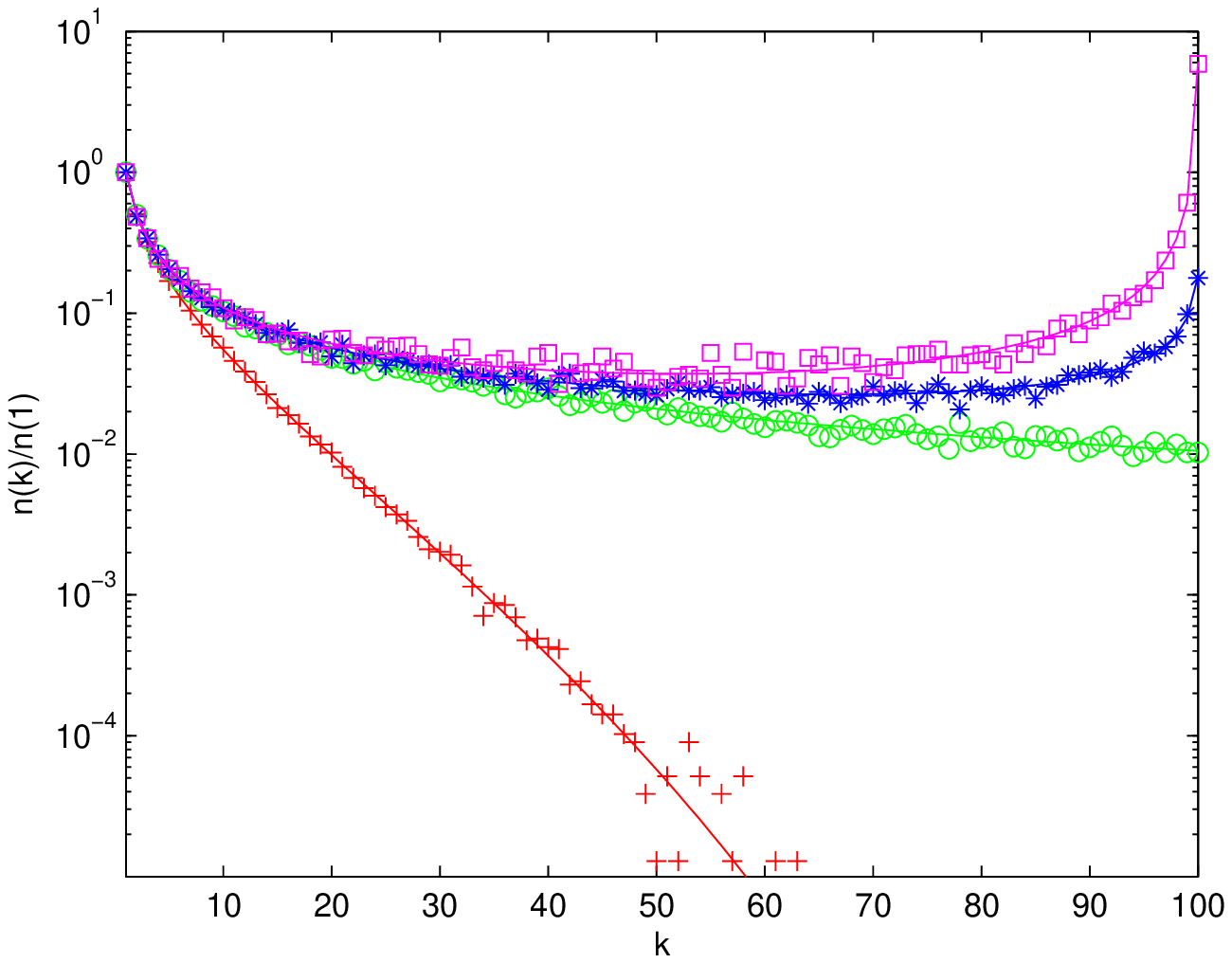}
\includegraphics[width=7.0cm]{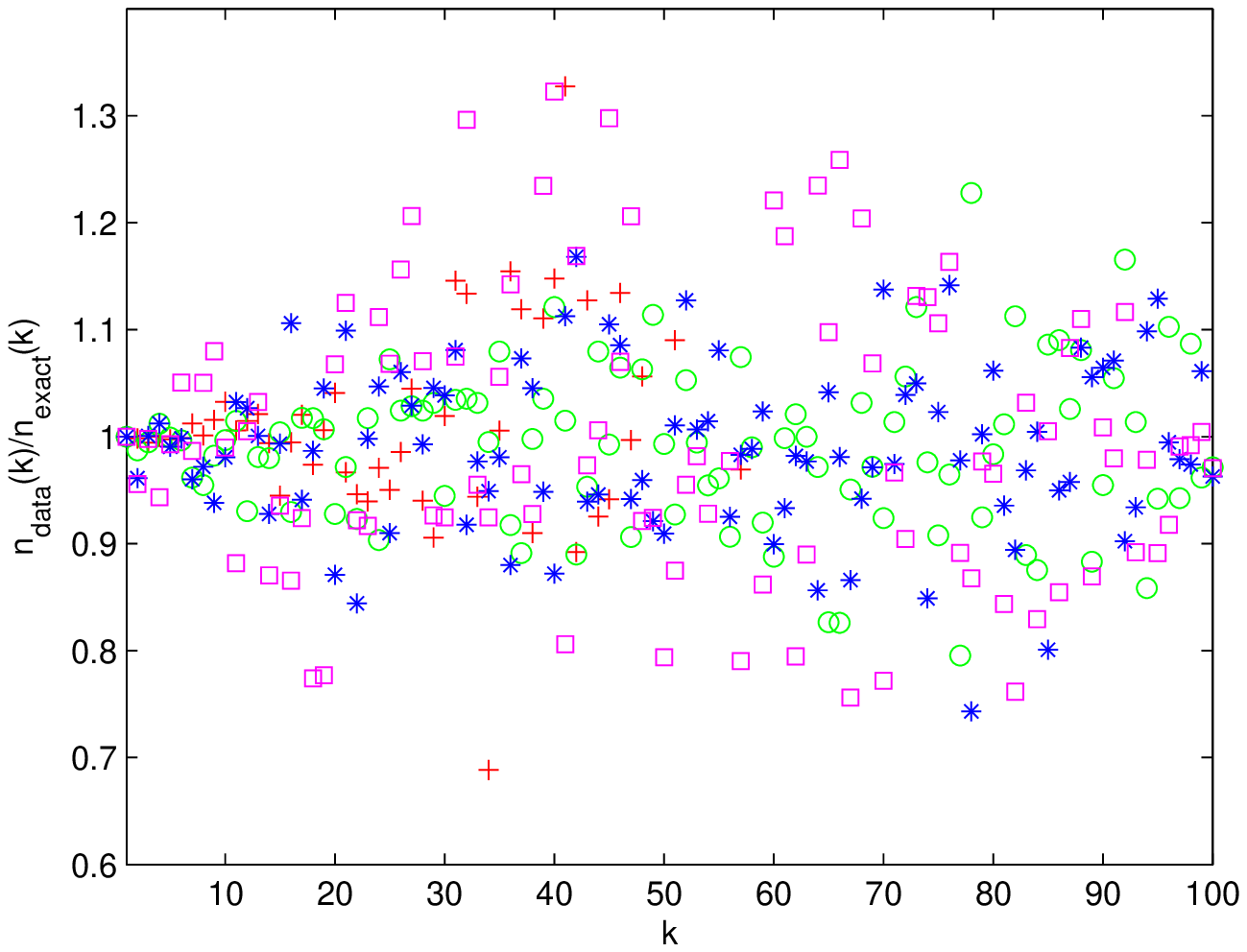}
\caption{Plots of the degree distribution normalised with respect to
$k=1$ and the fractional error of the data w.r.t.\ the exact
solution.  For $N=E=100$ and various $p_r=(1-p_p)=0.1$ (crosses),
$0.01$ (circles), $0.005$ (stars) and $0.001$ (squares), while lines
are the exact solutions. Measured after $10^5$ rewiring events,
averaged over $10^4$ runs. The size of fluctuations are clear from
the deviations about the exact solutions.} \label{fig:PLYDD}
\end{figure}

\begin{figure}
\includegraphics[width=7.0cm]{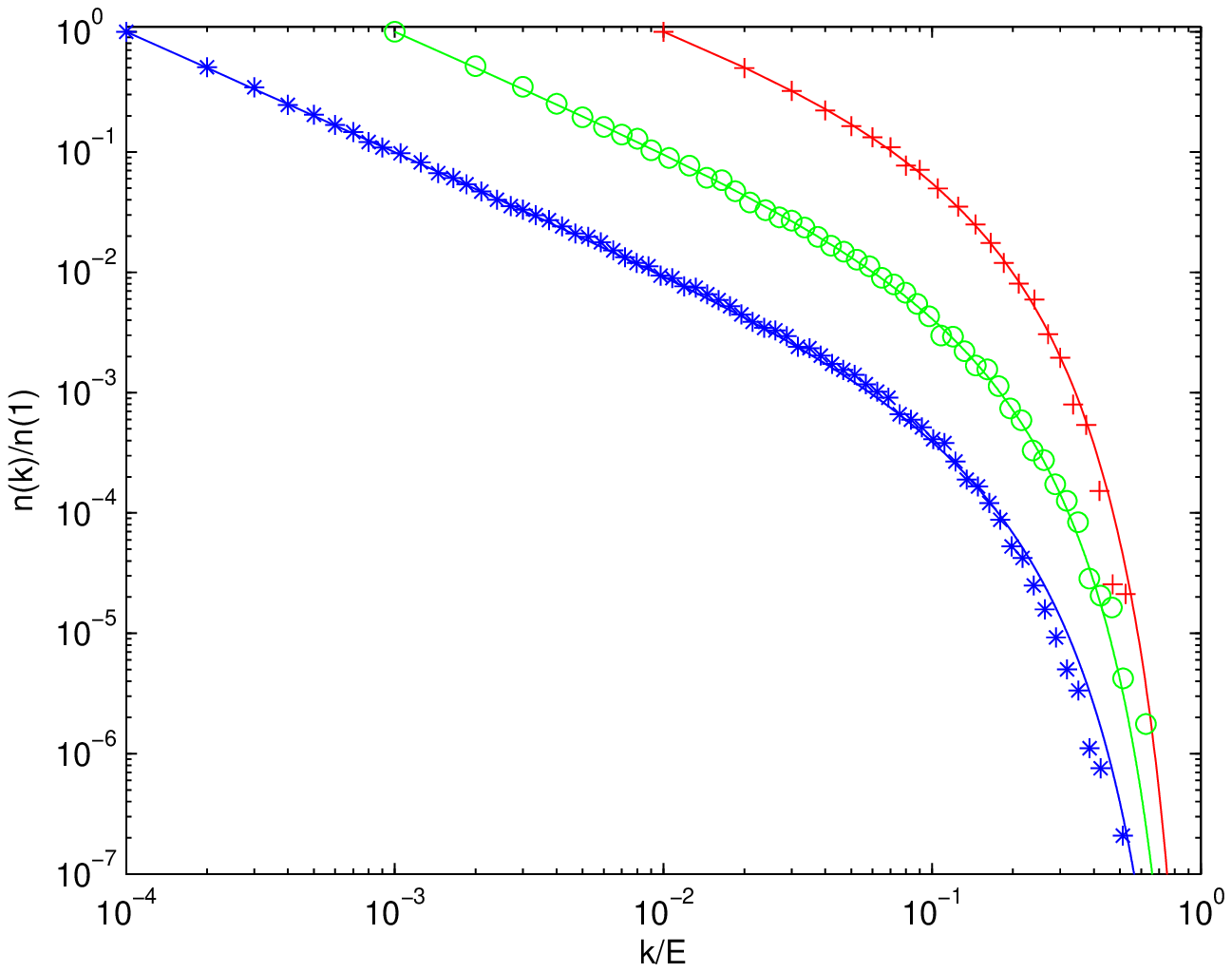}
\includegraphics[width=7.0cm]{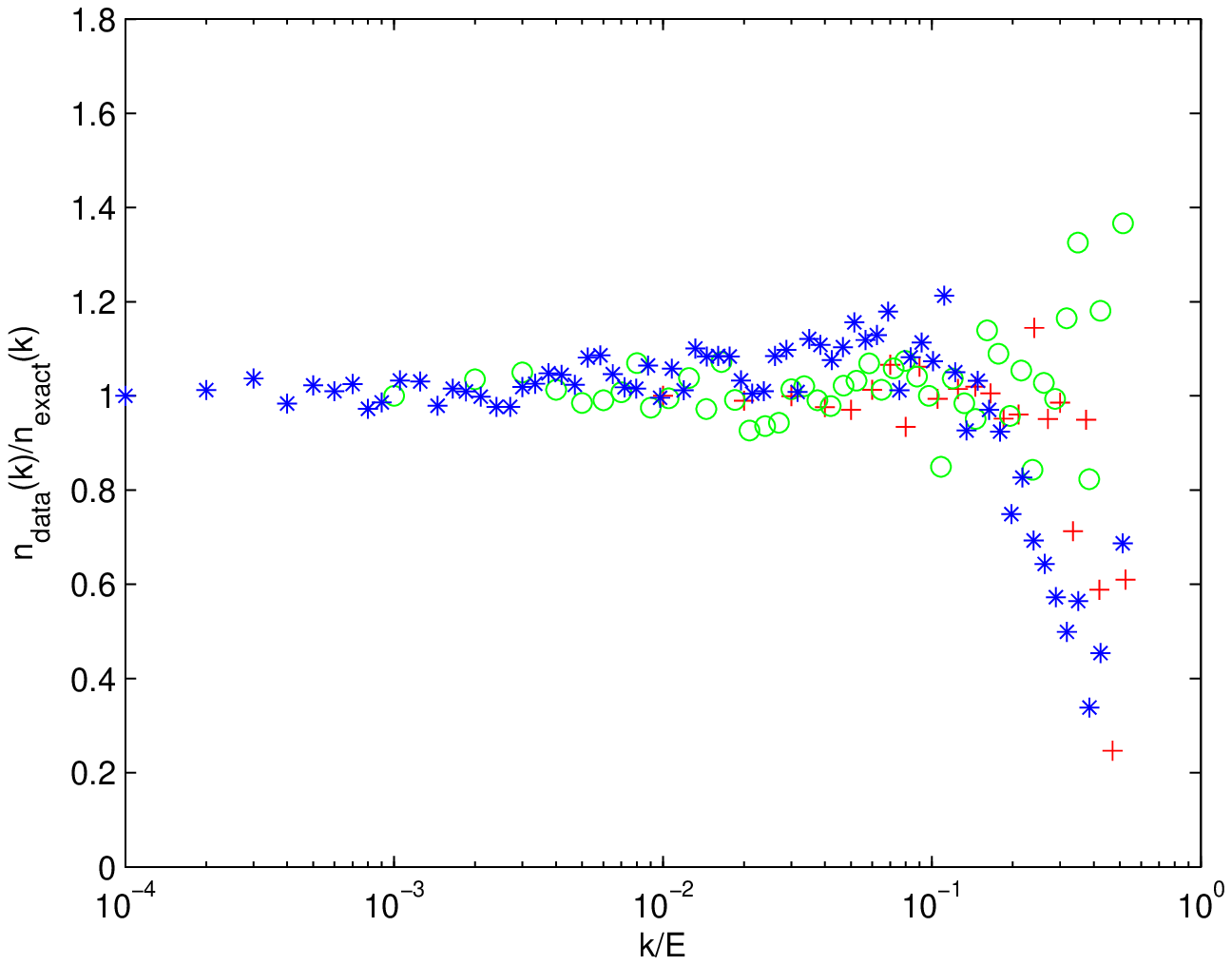}
\caption{The degree distribution normalised to $n(1)$ and the
fractional error w.r.t.\ the exact solution for $N=E$, $Ep_r=10.0$
and $p_r=10^{-2}$ (crosses), $10^{-3}$ (circles) and $10^{-4}$
(stars). Measured after $10^7$ rewiring events, averaged over $10^3$
runs. The tails of the distributions have few data points and so
they show larger fluctuations about the the mean field results, as
seen on the right. Despite this, for $p_r=10^{-4}$ there are clear
signs we have not quite reached equilibrium. Started with $n(k=1)=E$
but otherwise zero. } \label{fig:PLYDDEvary}
\end{figure}

These exact results for the degree distribution are for the mean
field equations.  These are only approximations but because in these
models there are no correlations between vertices, they should be
excellent approximations.  Simulations confirm this as
Figures~\ref{fig:PLYDD} and \ref{fig:PLYDDEvary} show.

\subsection*{The Generating Function}

Given the exact solution for the degree distribution \tref{nsimpsol}
its generating function $G(z)$, where
\beq
 G(z) := \sum_{k=0}^{E} n(k) z^k \; ,
\eeq
is found to be
\bea
 G(z) &=& n(0) F(\widetilde{K},-E;1+\widetilde{K}-E-\widetilde{E}; z)
 \\
 && \widetilde{K} = \frac{p_r}{p_p}\kav, \;
\widetilde{E} = \frac{p_r}{p_p}E .
\eea
Here $F$ is the Hypergeometric function. The average fraction of
sites of zero degree in the mean-field calculation is then exactly
\beq
 \frac{n(0)}{N}  =
 \frac{\Gamma(1-E-\Etilde)\Gamma(1+\Ktilde-\Etilde)}
      {\Gamma(1-E+\Ktilde-\Etilde)\Gamma(1-\Etilde)} .
\eeq

The $m$-th derivative of the generating function is
\beq
\frac{1}{G(1)} \left. \frac{d^mG(z)}{dz^m}\right|_{z=1} \!
 =
 \langle k(k-1) \ldots (k-m+1) \rangle
 =
 \frac{\Gamma(\widetilde{K}+m)\Gamma(-E+m)\Gamma(1-\widetilde{E}-m)}
      {\Gamma(\widetilde{K})\Gamma(-E)\Gamma(1-\widetilde{E})} .
\eeq
Knowing all the derivatives up to order $m$ gives all the moments
$\langle k^n \rangle$ up to that order. For case $m=1$ this provides
a consistency check on the parameter $\kav= E/N$ - the average
artifact degree.

\subsection*{New Artifact Addition}

\begin{figure}
\includegraphics[width=7.0cm]{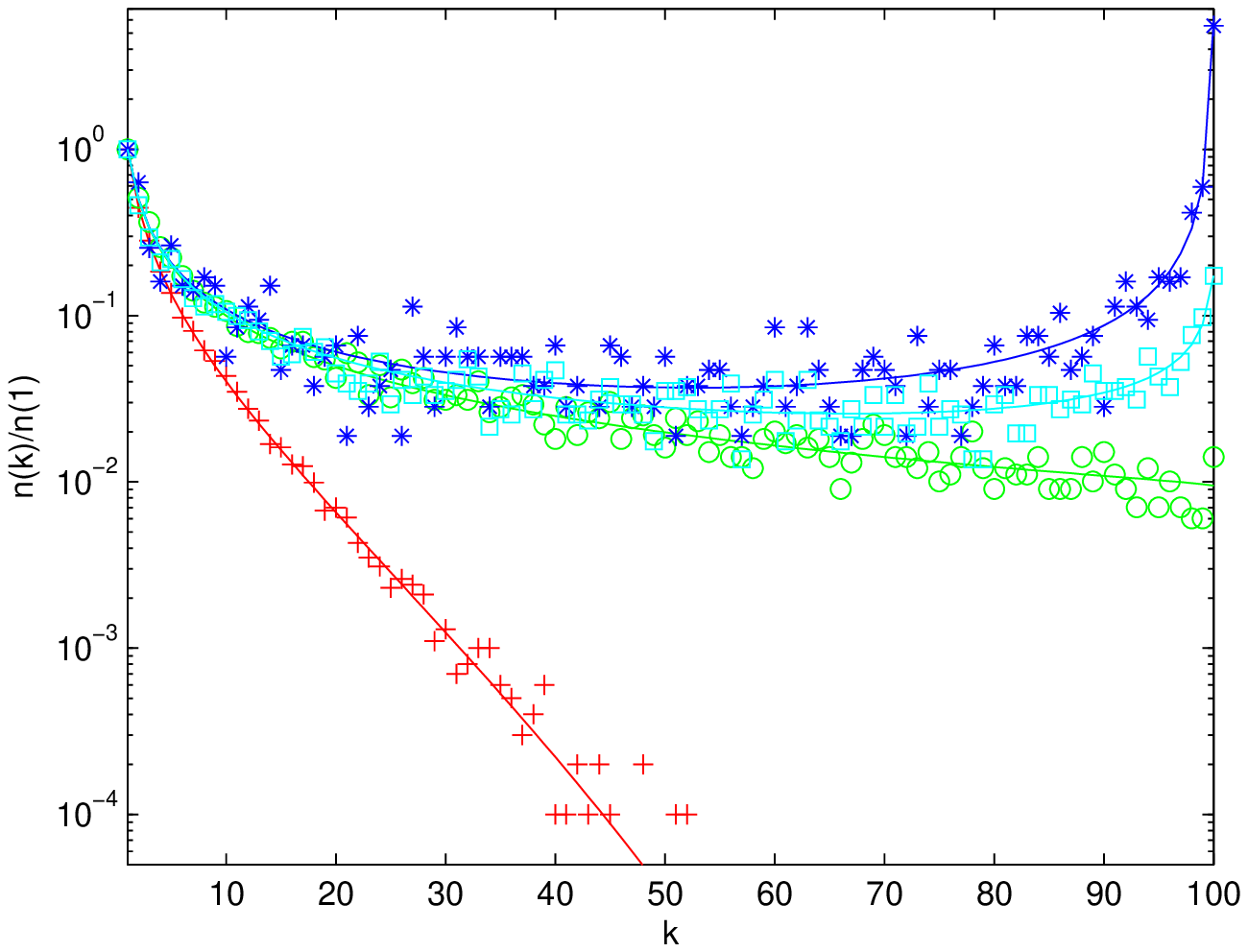}
\includegraphics[width=7.0cm]{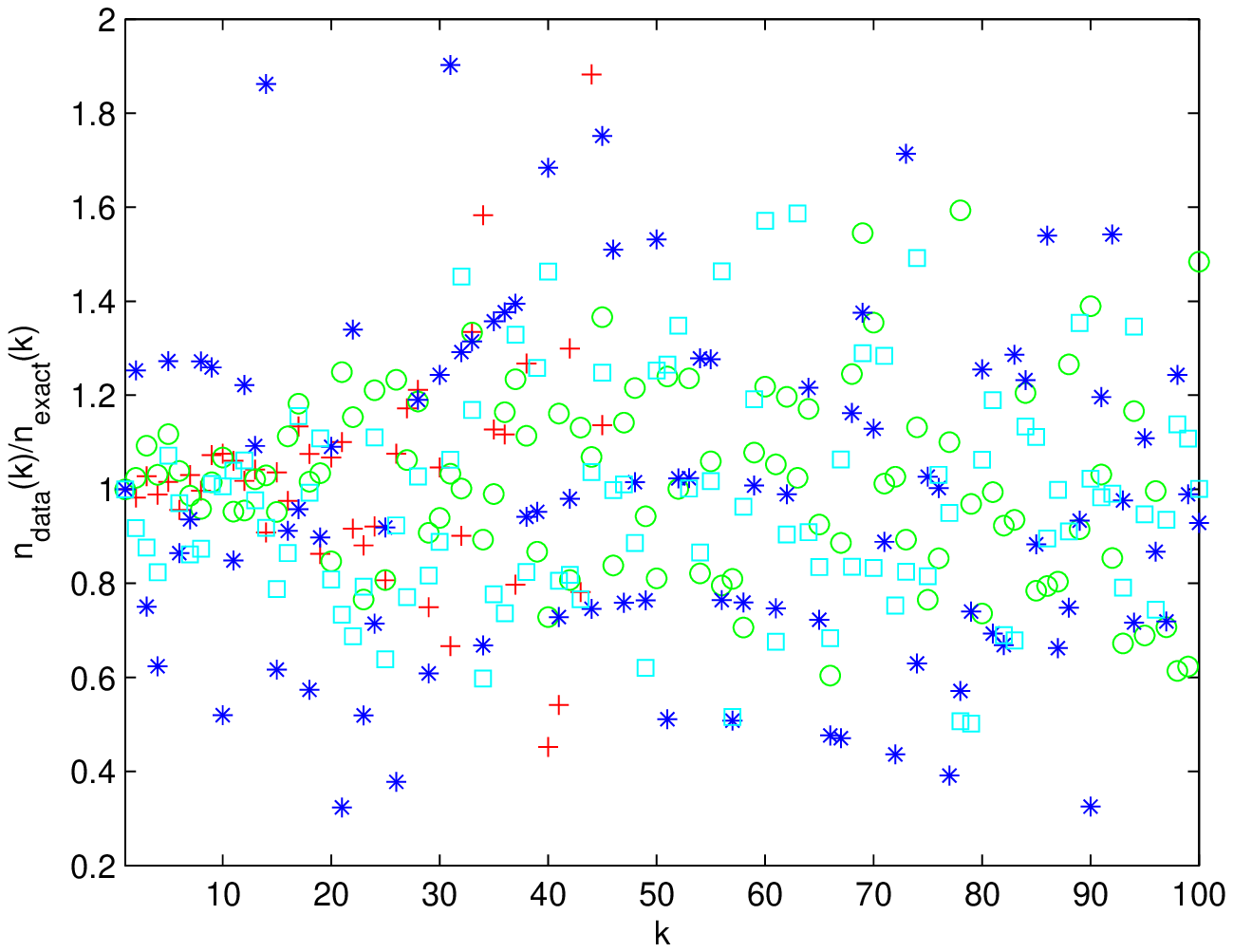}
\caption{Plots of the degree distribution normalised with respect to
$k=1$ and the fractional error of the data w.r.t.\ the exact
solution. For $E=100$ but with new artifacts added with probability
$p_r=0$, $\bar{p}=1-p_p$ where $\bar{p}=0.1$ (crosses), $0.01$
(circles), $0.005$ (squares) and $0.001$ (stars).  The lines are the
relevant exact mean field solutions. Measured after $10^5$ rewiring
events, and averaged over $10^4$ runs.} \label{fig:BHSDD}
\end{figure}

The cultural transmission models
\cite{Neiman95,BS03,HB03,BHS04,HBH04}, the gene evolution model of
\cite{KC64} and the model of family name distributions \cite{ZM01}
include another attachment process.  There a new artifact vertex is
added to the network with probability $\bar{p}=1-p_r-p_p$. The new
artifact receives the edge removed from an existing artifact on the
same time step. In this case the number of artifacts becomes
infinite so most artifacts have no edges.  Then the random
attachment becomes completely equivalent to this new process of
artifact addition. Thus the large $N$, zero $\kav$ limit of our
equations reproduces this case\footnote{Since $n(0)$ diverges this
must be excluded from discussions, but this is straightforward. An
alternative normalisation is needed, such as the number of `active'
artifacts $N_A= \sum_{k=1}^E n(k)$.}. The degree distribution for $k
\geq 1$ behaves exactly as above --- a simple inverse degree
power law cutoff by an exponential for $E(1-p_p) \gtrsim 1$ while
for $E(1-p_p) \lesssim 1$ $n(E) \approx 1$. In this model though,
when $p_r = p_{r, \mathrm{simple}}$ we have a degree distribution
which is an \emph{exact} inverse power law for the \emph{whole}
range of non-zero degrees. Our exact solutions to the mean field
equations again fits the data as Figure~\ref{fig:BHSDD} shows.

\subsection*{Equilibration Rate}

\begin{figure}
 \includegraphics[width=7cm]{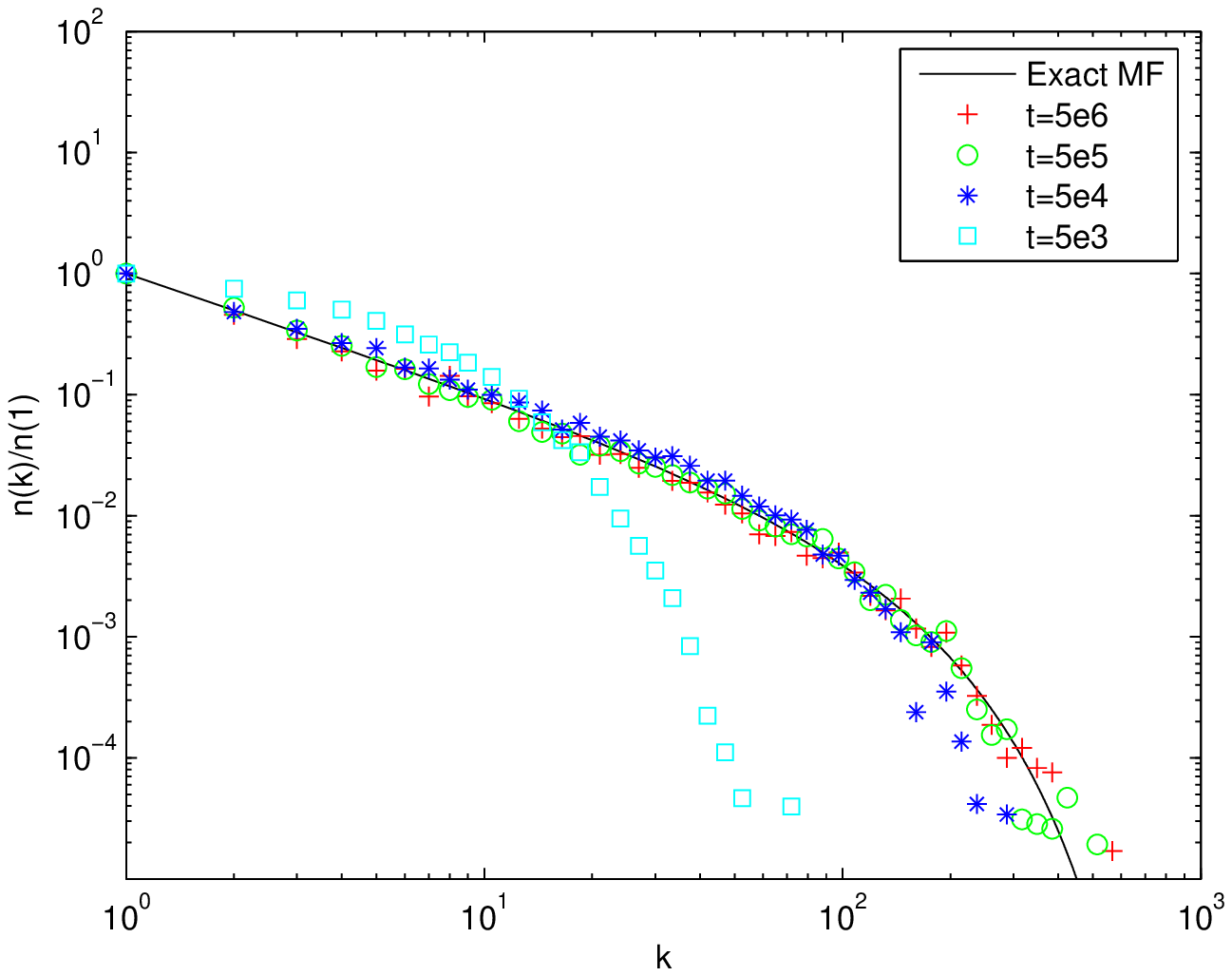}
 \includegraphics[width=7cm]{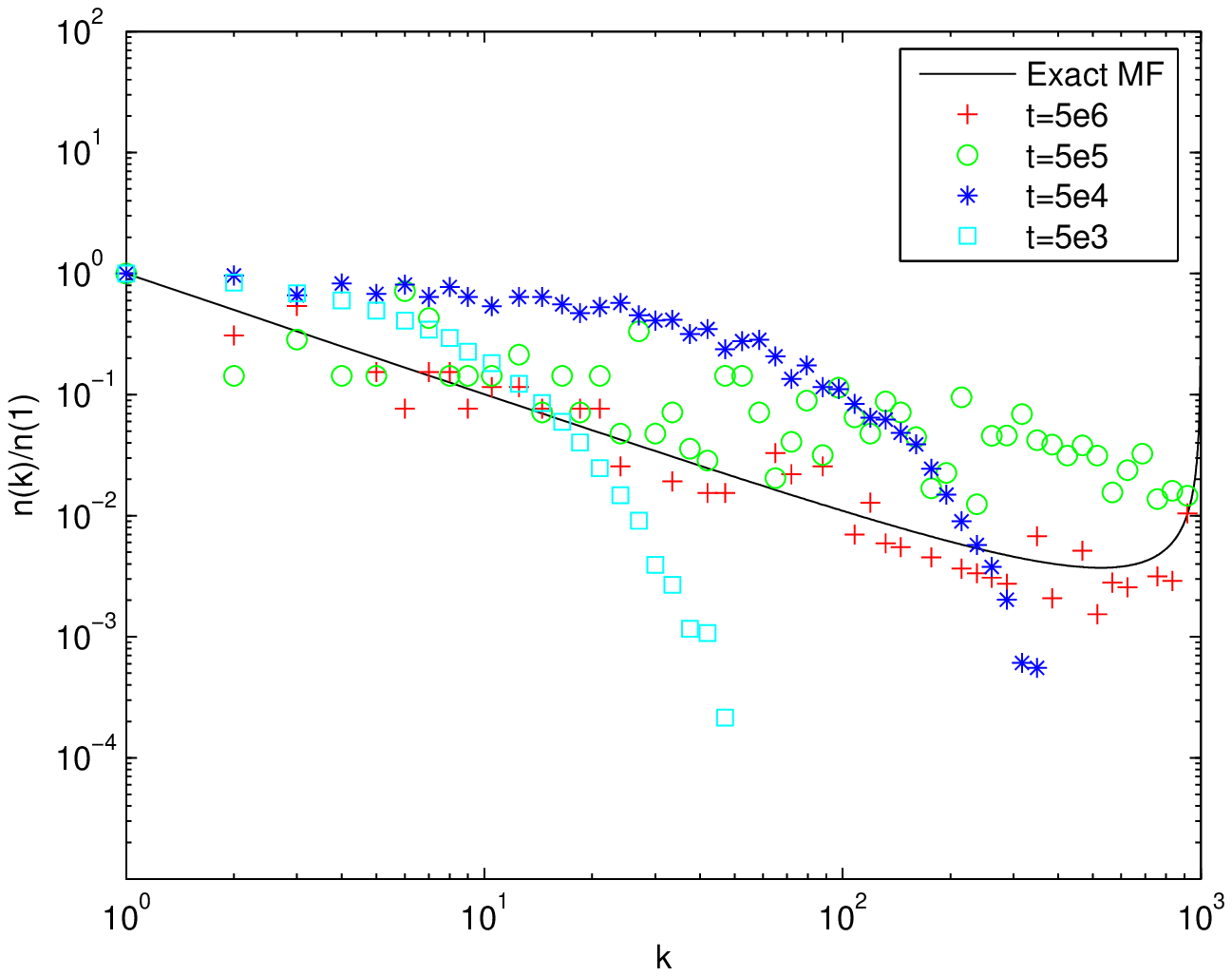}
 \caption{The degree distribution for $N=E=10^3$ for $(p_rE)=10$
(left) $(p_rE)=0.1$ (right), with $p_p=1-p_r$, $\bar{p}=0$, with the
degree distribution taken at a variety of times: $5\times 10^3$
(squares), $5\times 10^4$ (stars), $5\times 10^5$ (circles) and
$5\times 10^6$ (crosses). The initial configuration had each
artifact connected to one individual and results were averaged over
100 runs. With $(p_rE)=10$ we are close to equilibrium after $t=5
\times 10^3$ steps while this is only roughly true after $t=5 \times
10^5$ time steps for $(p_rE)=0.1$, consistent with our prediction
that the equilibration time scale is $\tau = E/p_r$.}
 \label{fBCMtvary}
\end{figure}

So far we have studied only the long time equilibrium distribution
of \tref{tdepsol} and therefore the first eigenvector of the matrix
$\Mmat$ of \tref{transmatrix} associated with eigenvalue 1. However
in Figure~\ref{fig:PLYDDEvary} there is clear evidence that the
system has not yet reached equilibrium despite the apparently large
number of rewiring events (each edge will have been rewired about
$10^4$ times and results were averaged over 100 runs).  This should be due to the second largest eigenvalue $\lambda_2$ in
\tref{tdepsol}.  We conjecture that this is of the form\footnote{We
have subsequently proved this conjecture \cite{EP06}.  In fact all
the eigenvectors and eigenvalues of $\Mmat$ of \tref{transmatrix}
have a distinctive pattern which we will report on elsewhere
\cite{EP06}.}
\beq
\lambda_2 = 1 - \frac{p_r}{E} \; .
\eeq
We have shown that this is always an eigenvalue of the transition
matrix $\Mmat$ \tref{transmatrix} and checked numerically that this
is indeed the second largest eigenvalue for $E<100$. The eigenvector
with this eigenvalue is
\begin{equation}
e^{(2)}_{E+1-k} =A^{(2)}
 \frac{\Gamma\left(k-\kav +1 \right)}
      {\Gamma\left(k-\kav \right)}
 \frac{\Gamma\left(k+\frac{p_r}{(1-p_r)}\kav \right)}
      {\Gamma\left( k+1 \right)}
 \frac{\Gamma\left(\frac{E}{(1-p_r)} -\frac{p_r}{(1-p_r)} \kav -k\right)}
      {\Gamma\left( E+1-k \right)  } \; .
\end{equation}
This means that the equilibration time scale --- the time taken for
contributions from eigenvectors $\evec^{(j)}$ ($j\geq 2$) to die
away --- is
\beq
 \tau = -\frac{1}{\ln(\lambda^{(2)})} \approx \frac{E}{p_r} =
 \frac{E^2}{(p_rE)}
 \label{taudef}
\eeq
The parameter which controls the shape of the distribution for most
examples is $(p_rE)$ and we expect to see the same shape independent
of $E$.  However what is noticable is that the rate of convergence
slows as $E^2$ as we increase $E$ for fixed $p_rE$.  This is visible
in Figure~\ref{fig:PLYDDEvary}. Figure~\ref{fBCMtvary} shows the
results are consistent with our prediction in \tref{taudef}.

\subsection*{Conclusions}

We have analysed the degree distribution in rewiring network models
and equivalent models which make no reference to a network. We have
shown that the mean field equations are different from the ones
considered in the literature. This makes little difference for
results quoted when $p_r E \gg 1$ but we have demonstrated that only
with the extra terms in \tref{neqngen} do we get the correct
solution for all values of $p_rE$.  Further we have found the second
eigenvalue and its eigenvector and thus deduced the rate of
convergence to the equilibrium solution.  This scales as
$E^2$ for fixed $p_rE$.

The literature suggests that probability distributions with a simple
inverse size form plus a large scale cut off, as found in these
models, are common. The real question is whether it is the copying
mechanism which leads to such distributions in practice?  It is
difficult to understand why in the real world individuals choose a
new artifact with a probability \emph{exactly} equal to the number
of times that artifact has been previously chosen, preferential
attachment. It is known that for growing networks deviations
from this law lead to deviations from power law distributions
\cite{KRL00}. Surely in the real world, many decisions would be
influenced by certain `leaders' in their fields and we are more
likely to copy their decision than that of other individuals?

In fact copying the choice of others, including that of certain
`leaders' may emerge naturally. Suppose our individuals were
connected to each other by a second network, a `contacts' network.
Individuals could use their contacts by copying the advice of a
friend or a friend of a friend as defined by the network of
contacts. This is equivalent to making a finite length random walk
on the graph of contacts. For growing networks this is known to be a
way that the structure of the graph can self-organise into a
scale-free form \cite{Vaz00,Vaz02,SK04,ES05} even if the random walk
is only one step long. In a similar way, for non-growing networks,
we are essentially making a one-step walk on the bipartite graph
between individuals and artifacts, regardless of any network between
the individuals.  Extrapolating the results of \cite{ES05} to the
non-growing case suggests that this should be sufficient to generate
an effective attachment probability of the form \tref{PiRPiAsimple}.
Put simply we expect that whatever we do, the probability of
arriving at one artifact at the end of a random walk is going to be
dominated by the number of routes into that artifact, i.e.\ its
degree.

Such an example may be seen in the the model of the minority game by
Anghel et al.\ \cite{ATBK04}.  Their individuals are connected by a
random graph and at each time step  an individual copies the best
strategy (the artifact in this case) from amongst the strategies
of their neighbouring contacts. The individuals do not choose a
random neighbour's artifact but the `best' artifact.  However if the
meaning of best is always changing, as it may be in the Minority
Game or in many examples of fashion, this best choice may be
effectively a random neighbour choice and hence be statistically
equivalent to the simple copying used in our models. Thus even if it
appears that the population is influenced by wise men or fashion
leaders, provided there is little substance to their choices then it
may well be equivalent to simple copying of one person's choice. It
should come as no surprise that the results for the artifact degree
distribution in \cite{ATBK04}, the popularity of the most popular
strategies, follows a simple inverse power law with a large degree
cutoff, exactly as the simple copying model would give.

Finally one might ask if it is important to get the right
classification of artifacts to see the distribution.  What if people
make choices based on one classification but we measure on another?
Do people choose a specific breed of dog as registered by the dog
breeders association of their country, or do they really choose
between small and large dogs, short or long haired dogs
\cite{HBH04}?  The classification of pottery in archaeology is one
imposed on the record by modern archaeologists. The answer ought to
be that the classification should not be important and it is a
scaling property of the model and its solutions that this is so.

Suppose we randomly paired all the artifacts but deemed the
fundamental process to be based still on the choice of the original
artifacts and their degree. The choice of edge to remove is
unchanged while preferential attachment to the underlying individual
artifacts leads to effective preferential attachment to the artifact
pairs. The probability of an innovation event ($p_r$ or $\bar{p}$)
is unchanged but the probability of choosing a random artifact pair
is double that of choosing a single artifact. However that reflects
the fact that the number of artifacts pairs $N_2$ is half the
original number of artifacts. Thus we see that we require that $N
\rightarrow N_2 = (N/2)$ but we keep all other parameters the same.
In particular the linear nature of preferential removal and
attachment to the artifact pairs means that the form of both removal
and attachment is unchanged.  Overall we have exactly the same
equations for the artifact pair degree distribution $n_2(k)$ as we
did for the original artifacts. Thus the distribution $n(k)$ of
\tref{nsimpsol} is of the same form with $N\rightarrow (N/2)$ being
the only change required. However we have seen that for $Ep_r
\gtrsim 1$ the shape can be parameterised in terms of a power
$\gamma$ \tref{gammasimp} and an exponential cutoff $\zeta$
\tref{zetasol}.  The latter is unchanged and while the power does
change a little,  we have argued that if $\gamma$ can be measured,
it is likely to be indistinguishable from one in any real data set.
So apart from the overall normalisation, the distribution of
artifact choice is essentially independent of how we choose to
classify the artifacts. This stability against the classification of
the artifact types is an important feature of the copying models
considered here.

TSE would like to thank H.Morgan and W.Swanell for useful discussions.



\tpre{
 \newpage
 \section*{Supplementary Material}
 This plot is not in the published version.
 \begin{figure}[hbt]
 \includegraphics[width=7cm]{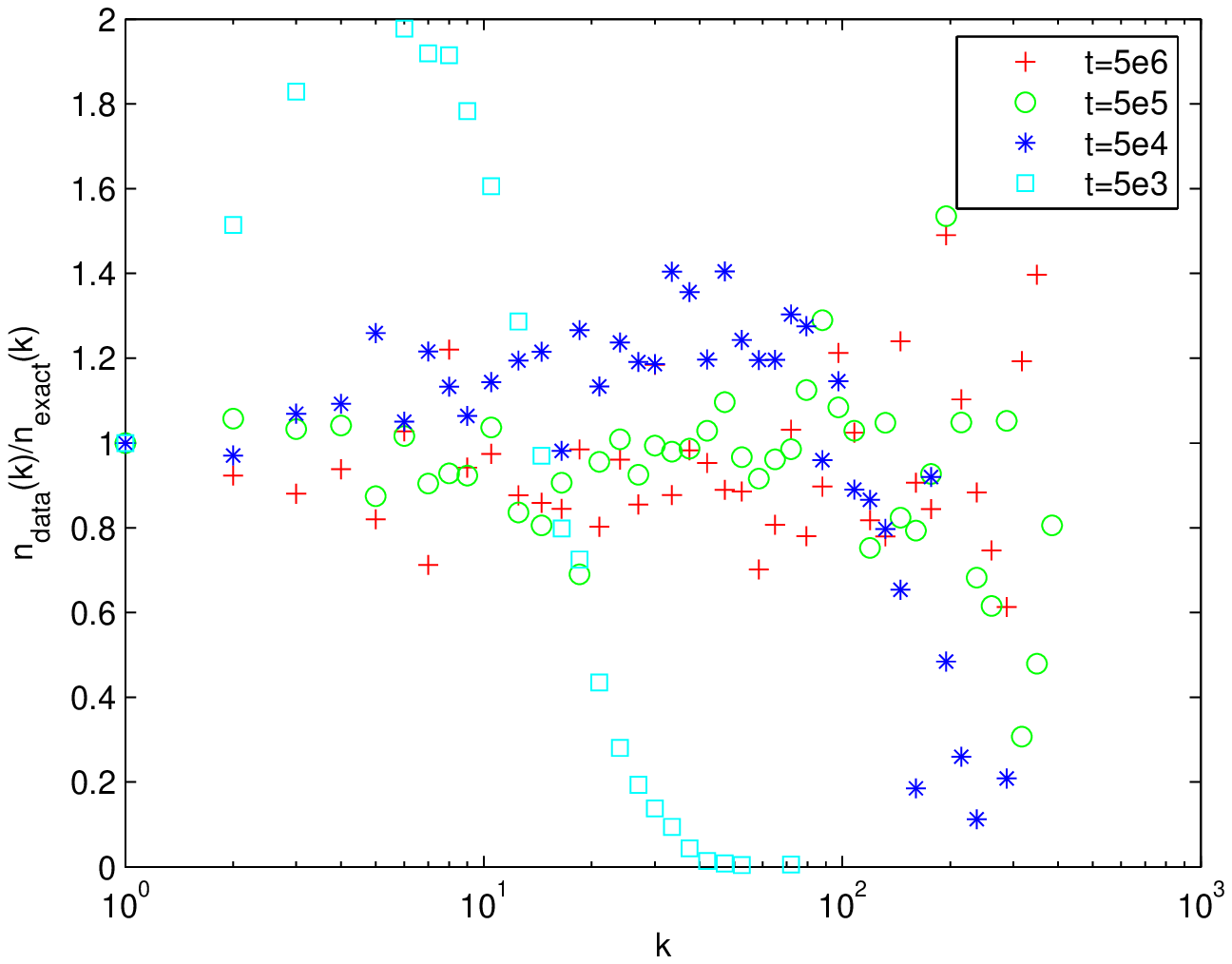}
 \includegraphics[width=7cm]{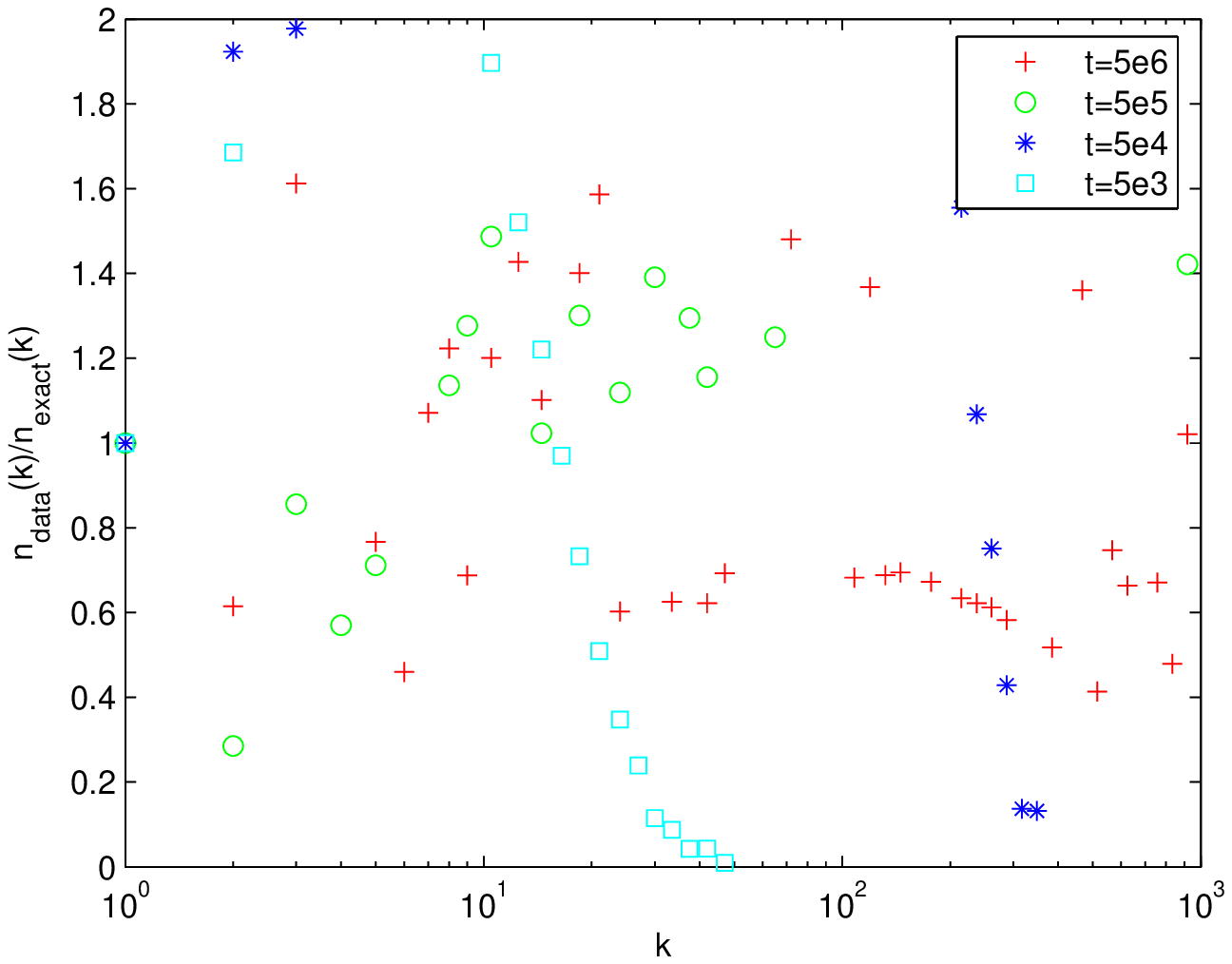}
 \caption{Plots of the fractional error of the data of Fig.\ \ref{fBCMtvary}
w.r.t.\ the exact solution for $(p_rE)=10$ (left) $(p_rE)=0.1$
(right).  With $N=E=10^3$, $p_p=1-p_r$, $\bar{p}=0$, with the degree
distribution taken at a variety of times: $5\times 10^3$ (squares),
$5\times 10^4$ (stars), $5\times 10^5$ (circles) and $5\times 10^6$
(crosses). The initial configuration had each artifact connected to
one individual and results were averaged over 100 runs. With
$(p_rE)=10$ we are close to equilibrium after $t=5 \times 10^3$
steps while this is only roughly true after $t=5 \times 10^5$ time
steps for $(p_rE)=0.1$, consistent with our prediction that the
equilibration time scale is $\tau = E/p_r$.}
 \label{fBCMtvaryFracErr}
\end{figure}
}

\end{document}